\documentclass[manuscript]{jkas}

\def\year{2017} % publication year
\def\month{May} % publication month
\def\volume{00} % journal volume
\def\issue{0} % journal issue
\def\beginpage{1} % first page of article
\def\endpage{99} % last page of article
\def\received{January 30, 2017} % date paper was received by JKAS
\def\accepted{February 28, 2017} % date of acceptance
\date{Received \received; accepted \accepted}

\usepackage{bm}
\usepackage{color}%, ulem}

\newcommand{\ba}{\begin{eqnarray}}
\newcommand{\ea}{\end{eqnarray}}
\newcommand{\non}{\nonumber}
\newcommand{\T}{\mathrm{T}}

\newcommand{\apj}{ApJ}
\newcommand{\aap}{A\&A}
\newcommand{\aj}{AJ}
\newcommand{\pasj}{PASJ}
\newcommand{\araa}{ARA\&A}
\newcommand{\aaps}{A\&AS}
\newcommand{\mnras}{MNRAS}

\title{
Time Variations of the Radial Velocity of H$_2$O Masers in the Semi-regular Variable R Crt
}

\author[1,*]{Hiroshi~Sudou}
\author[1,2,*]{Motoki~Shiga}
\author[3]{Toshihiro~Omodaka}
\author[3]{Chihiro~Nakai}
\author[1]{Kazuki~Ueda}
\author[1]{Hiroshi~Takaba}
\affil[1]{Faculty of Engineering, Gifu University, 1-1 Yanagido, Gifu, Gifu 501-1193, Japan; \email{sudou@gifu-u.ac.jp}}
\affil[2]{Precursory Research for Embryonic Science and Technology (PRESTO), Japan Science and Technology Agency (JST), 4-1-8, Honcho, Kawaguchi, Saitama 332-0012, Japan; \email{shiga\_m@gifu-u.ac.jp}}
\affil[3]{Faculty of Science, Kagoshima University, 21-24 Korimoto, Kagoshima, Kagoshima 890-8580, Japan; \email{omodaka@sci.kagoshima-u.ac.jp}}
\affil[*]{The first two authors (HS and MS) equally contributed to this paper.}
\def\corrauthor{
H. Sudou and M. Shiga
}

\def\runningauthor{
}

\def\runningtitle{
Monitoring Water Masers in R Crt
}

\def\keywords{
masers --- stars: individual (R Crateris) --- stars: late type --- stars: AGB and post-AGB --- stars: outflows
}

\def\abstracttext{
H$_2$O maser emission {at 22 GHz} in the circumstellar envelope is one of the good tracers of detailed physics and kinematics in the mass loss process of asymptotic giant branch  stars.
Long-term monitoring of an H$_2$O maser spectrum with high time resolution enables us to clarify acceleration processes of the expanding shell in the stellar atmosphere. 
We monitored the H$_2$O maser emission of the semi-regular variable R Crt with the Kagoshima 6-m telescope,
and  obtained a large data set of over 180 maser spectra during 1.3 years with an observational span of a few days.
Based on an automatic peak detection method based on a least-squares fitting with a Gaussian basis function model, 
we exhaustively detected peaks as significant velocity components with the radial velocity on a 0.1 km s$^{-1}$ scale.
This analysis result shows that the radial velocity of red-shifted and blue-shifted components exhibits a change between acceleration and deceleration on the time scale of a few hundred days.
These velocity variations are likely to correlate with the intensity variations, in particular during flaring state of  H$_2$O  masers.
It seems reasonable to consider that the velocity variation of the maser source is caused by the shock propagation in the envelope due to stellar pulsation. 
However, it is difficult to explain the relationship between the velocity variation and the intensity variation only 
from shock propagation effects. 
{  We found that a time delay of the integrated maser intensity with respect to the optical light curve is about 150 days. }
}

\begin{document}
\jkashead %% set title, authors, abstract, etc.

%%%%%%%%%%%%%%%%%%%%%%%%%%%%%%%%%%%%%%%%%%%%%%%%%%%%%%%%%%%%%%%%%%%%%
%%% BEGIN MAIN TEXT HERE %%%%%%%%%%%%%%%%%%%%%%%%%%%%%%%%%%%%%%%%%%%%
%%%%%%%%%%%%%%%%%%%%%%%%%%%%%%%%%%%%%%%%%%%%%%%%%%%%%%%%%%%%%%%%%%%%%

\section{Introduction}

Understanding the mass loss process of the circumstellar envelope of asymptotic giant branch (AGB) stars is one of the long-standing issues both in the evolution of low mass stars
and in the chemical evolution of matter in the Universe. 
It is generally accepted that the mass loss process in AGB stars is caused by radiation pressure on dust grains which launches the materials outwards (H\"{o}fner 2008, 2009). 
Recent high-resolution Atacama Large Millimeter/submillimeter Array (ALMA) observations have shown anisotropy of dust distribution such as clumpy structure (O'Gorman et al. 2014) and spiral structure (Decin et al. 2014). 
These facts indicate that the dust distribution is very complex and  not always isotropic and homogeneous. 

AGB stars often show the maser emission from SiO, H$_2$O, and OH molecules.
Recent simultaneous observations of multi maser lines with the Korean VLBI Network telescoes showed the statistical
properties of late evolutionary stages from AGB to post-AGB stars  (Kim et al. 2014; Yoon et al. 2014; Kim et al. 2016),
in particular, the H$_2$O maser emission at 22 GHz associating with the outer region of the dust shell is strongly affected 
by expanding motion of the envelopes and shock waves generated by the stellar pulsation.
Thus, observations of the H$_2$O masers can be useful to trace the detailed motion of the materials in the envelope of mass losing stars,
which was also revealed by some homogeneous databases provided by sensitive radio telescopes (e.g., Medicina 32m, Comoretto et al. 1990, Valdettaro et al. 2001).

The H$_2$O maser spectra show many velocity peaks, typically double peaks, due to the Doppler shift which reflects the radial speed of the maser sources. 
By tracing the velocity in the H$_2$O maser spectra, we will be able to obtain information of the acceleration process in the envelope.
Very-long-baseline interferometry (VLBI) mapping monitoring revealed that in some giant stars, the H$_2$O maser is distributed in the circumstellar shells with radii of several tens of AU.
It has also indicated that it is accelerated up to $\sim$10 km s$^{-1}$ by radiation pressure from the surrounding dust shell (Elizur 1992), 
and acceleration at the 0.1 km s$^{-1}$ yr$^{-1}$ level (e.g., Richards \& Yates 1998). 
%%%%%Both single-dish and VLBI monitoring are regarded as a complementary method each other, because it is of great value to have very frequent and long term data sets of many stars emitting the H$_2$O maser. 

The semi-regular variable stars are known as a type of AGB stars and sometimes as H$_2$O maser emitters. 
They are good candidates for searching for the anisotropic property of the mass loss in the AGB phase, because they exhibit complex light variations. 
Although the origin of the irregularities is not clear, possible explanations include such as multi-periodic pulsations, non-radial oscillations, or the presence of a convection shell (Hinkle et al. 1997, Lebzelter et al. 2000 and references therein). 

R Crt is a semi-regular variable classified as SRb of M7 spectral type (Kholopov et al. 1987). 
The $V$ band magnitude varies with a pulsation period of 160 days indicated from the American Association of Variable Star Observers (AAVSO) data base. 
The H$_2$O masers in R Crt show a double peak separated by 7 km s$^{-1}$, indicating that R Crt has typical maser spectral properties of AGB stars with the expanding shell (Takaba et al. 1994). 
In particular, the H$_2$O masers in R Crt have been examined intensively in order to study its velocity structure. 
VLBI monitoring during six months revealed the 3D velocity structure indicating the possible bipolar structure of the H$_2$O maser region with velocity ranging from 4.3 to 7.3 km s$^{-1}$ (Ishitsuka et al. 2001, hereafter I01). 
Further, single-dish monitoring during 3 years showed complex velocity variability in the H$_2$O maser spectrum on the 0.1 km s$^{-1}$ scale (Shintani et al. 2008).  
They also found switching between acceleration and deceleration in some velocity components, which can be  evidence of periodic events and stellar pulsation.
%, but it is unclear whether or not it is periodic and relates to stellar pulsation.
Although these observations advanced understanding of the dynamics of the envelope in R Crt, it is necessary to carry out more massive monitoring observation to understand the velocity variation in detail.  %However, manually performing this task is difficult to avoid a lot of analysis cost (time) and possible individual differences, because such frequent monitoring will produce a lot of datasets daily. 
In order to search for the complex velocity field in the circumstellar envelope in R Crt, we carried out frequent (once a few day) spectral monitoring of the H$_2$O maser with velocity resolution on the 0.1 km s$^{-1}$ scale.  
Tracking velocity components by detecting all peaks in the maser spectra enables us to accurately monitor the activity of the circumstellar envelope in detail. We have exhaustively detected such all peaks using our developed code of an automatic peak detection based on a Gaussian basis function model.

\section{Observations and Data Reductions}

A time series of the H$_2$O maser spectra of R Crt at 22 GHz was observed during 1.3 years from January 2008 to March 2009, 
by using the 6-m telescope at Kagoshima, Japan (Omodaka et al. 1994). 
In total, we have obtained 187 maser spectra. It corresponds to the observing frequency of once a few days on average. 
The digital spectrometer installed at the telescope has 160,000 channels, which provides 0.05 km s$^{-1}$ per a channel, 
and the spectral range covered 64 MHz (830 km s$^{-1}$). 
The typical observation error of the velocity of the 6-m telescope is estimated to be 0.1 km s$^{-1}$,
mainly limited by the accuracy of the velocity calibration for the Earth rotation parameters.
The standard on-off beam-switching technique was used for the observation of the source. 
%%%%%The integration time of each observation was xx min in typical. 
%%%\red{  were typically between 200 -- 300 K (?)}, 
The system noise temperatures were measured at each observation and were used to correct the antenna temperature 
for atmospheric attenuation and changes in antenna gain as a function of the elevation angle.
The typical error of the amplitude is estimated to be 10 \% in our R-sky calibration system of the 6-m telescope. 

%The systemic velocity is 10.8 km s$^{-1}$ based on the results from CO observation profile (Bowers 1992).
%We found many H$_2$O maser peaks in the velocity rage of 2 -- 18 km s$^{-1}$. The spectrum usually presents three main components at 5, 10, and 15 km s$^{-1}$, corresponding to blue-shifted, systemic, and red-shifted components, respectively.   

Figure 1(a) shows an example of the H$_2$O spectrum observed on January 13, 2008. 
This figure shows that there are several peaks in this spectrum.
We found three main components; the components peaked around 10.8 km s$^{-1}$, which is the systemic velocity from the result of the CO observation profile (Bowers 1992),
and those peaked around 5 and 15 km s$^{-1}$, which are blue-shifted and red-shifted components, respectively.
The velocity distribution of these components is likely to be interpreted as a typical result from the expanding shell (I01). 

%Large monitoring increases the data size rapidly and it makes manual data analysis difficult to carry out because of an increase of the analysis time and possible individual differences among data analysts.
Frequent monitoring over a long time period increases the size of spectral datasets to be analyzed and then it makes manual analysis difficult 
because of the data analysis cost and human errors. Against this problem, automatic Gaussian deconvolution (AGD) algorithm, 
which is a fitting method with a Gaussian basis function model, was proposed in the massive 21-cm absorption survey (Lindner et al. 2015). 
The advantage of this approach is to automatically and exhaustively detect peaked components over entire observed spectra. 
Thus, this approach can drastically reduce both data analysis cost and human errors.

The peak detection by AGD algorithm is implemented based on a least-squares fitting with the basis function model. 
The algorithm initializes function parameters using several conditions with the higher order derivatives and then optimizes function parameters 
by a gradient descend method.  However this algorithm can fail to reach to good local optima to detect peaks correctly because AGD algorithm uses 
the initialized parameters chosen by the proposed condition but it is still affected by observation noise. To avoid this problem, 
we took an approach that repeatedly and randomly initializes parameters presumed as peaked and then optimized parameters from all initialized parameters, 
and finally chooses the best fitting result minimizing the mean squared error {  (see Appendix in detail)}. 
{  After the optimization, less significant basis functions whose intensities are smaller than the threshold   {$w_{\rm thre}$ } are removed from further analysis processes. 
Figure 1(b) shows an example of our fitting result by the Gaussian basis function model with the threshold   {$w_{\rm thre}$ = 0.3} { Jy/beam} and 
center positions of peaked components  (basis functions) are indicated by symbol $\triangle$. }
%%%\red {  The typical noise level in our spectrum is 0.1 Jy/beam, i.e., {$w_{\rm noise}\sim3 \sigma$}}
%%%\red {(see Appendix in detail).}
{  This selection allows us to remove weak and short-lived components which are hard to be identified at different observation dates. 
This figure shows that bright four peaks were detected by our method. }
%%%The detailed procedure is described in the appendix.

\section{Results}

\subsection{Time series of the integrated intensity}
{  Figure 2 shows the integrated total intensity for the all detected components. 
The variability curves presented two clear peaks around JD = 2454560 and 2454870 during our monitoring period,
and the separation of these peaks is about 300 days.}

\subsection{Time series of the peak velocity of each maser component}

Figure 3 shows the time variations of the radial velocity of the detected peaks, ranging from 0 to 20 km s$^{-1}$. 
Some velocity components seem to have a lifetime of more than one year.
These results are generally in agreement with the previous work, by which involved monitoring with the VLBI Exploration of Radio Astronomy (VERA) 25-m telescope in Iriki (Shintani et al. 2008). 
{  On the other hand, other components with a shorter lifetime seem to be more variable compared with blue- and red-shifted components.
These highly variable components might be associated with irregular fluctuations in the envelope indicated by semiregular optical variations of the optical
light curve in R Crt \citep{rud10}.
In our present data, it is difficult to find the relationship between the velocity variation and the optical variation, 
because our monitoring is not so sensitive to trace the variation of these weak components correctly.
}

Figure 4 shows details of the time variations of blue-shifted, systemic, and red-shifted components, which lived for more than a year. 
In the blue-shifted components,  we can find the velocity variations ranging $\pm$ 0.2 km s$^{-1}$ (figure 4(a)). 
The timescale of the variation is estimated very roughly from peak-to-peak measurements to be 300 days. 
{   This is very similar to the timescale of the variation of the integrated intensity shown in Figure 2. }
The red-shifted component is also likely to show the similar velocity variation as that of the blue-shifted component. 
Its timescale is estimated to be roughly 200 days. 
The timescale obtained in the flux monitoring of the optical and radio waves has been reported the periods of the AAVSO's optical light curve of 160 days and of the time variations of the OH masers in R Crt of 227 or 560 days (Etoka et al. 2001). 
It is unclear whether or not the timescale of the velocity variation is related to the optical and radio flux variations from our limited results.
%%%This timescale is similar to the periods of the optical or OH maser variations, in contrast with the blue-shifted component. 

The largest velocity shift can be seen in the systemic component. 
According to the simple expanding shell model, the shift of the radial velocity of the systemic components is expected to be very small,
because they are expected to move to perpendicular to the line of sight.  
Similar behavior was reported in the previous paper of VLBI monitoring of R Crt (I01). 
They found that the largest velocity shift of $\sim -4$ km s$^{-1}$ yr$^{-1}$ was found for the components near the systemic velocity. 
Perhaps it is related to the bipolar flow in the H$_2$O maser shell suggested from the proper motion of the maser spots, or to the presence of non-radial oscillation. 

We also show the time variation of the velocity dispersion of each component in Figure 4. 
{  The clear systematic variation of the velocity dispersion of the blue-shifted and red-shifted components cannot be found before JD = 2456500.
Then the velocity dispersion becomes narrower from JD = 2454650 to 2454750, after that it becomes slightly wider.  
Although this tendency might show anti-correlation with the intensity variations, further monitoring should be needed to investigate it.}
%%%after that it become slightly wider. 
%%%Since this tendency is likely to show anti-correlation with the intensity variations, 
%%%it is likely to suppose that enhancement of the dispersion are caused by the low noise-to-signal ratio.

%%%On the other hand, the velocity width of the systemic component is relatively wider than those of the other components,
%%%and does not show the clear change after JD = 2454700.  

%%%The acceleration was estimated from the least mean square method by using the velocity change in a month. 
%%%we can find that their behavior looks like possible oscillating variations.  

\subsection{Time series of the peak intensity of each maser component}

We show the light curve of R Crt obtained by the All Sky Automated Survey (ASAS, Pojmanski 1997) in Figure 4(d).
Although it is remarkable that the big flare of the blue-shifted component can be seen near JD=2454550, corresponding to $\sim 9$ times the intensity just before the flare, 
there are no strong indications for the ASAS's optical light curve.
The red-shifted component also showed the similar but smaller flares near both 2454550 and 2454650 days. 
{  This fact indicates that the blue-shifted component is more active and variable compared with the red-shifted component during our monitoring. 
This activity might relate to the fact that the velocity variations for the blue-shifted component are almost twice as large as those for the red-shifted component (see \S 3.2).
}

It has been known that intensity variations of masers are difficult to interpret because it relates to the pumping mechanism such as heading and compression of the gas by shock (G$\rm \acute o$mez Balboa \& Lepine 1986).

\subsection{Correlation between the peak velocity and the peak intensity}

Figure 5 shows the relationship between the radial velocity and the intensity of each component.  
Interestingly, it looks that the velocity variations are correlated with the intensity variations, that is, the expanding velocity increases (acceleration) with increasing the intensity. 
Pearson correlation coefficient (cc) is $-$0.34 with p-value of $2.2\times 10^{-6}$ for the blue-shifted shown in Figure 5 (a) and is 0.38 with p-value of $5.4\times 10^{-7}$ for red-shifted components in Figure 5(c).
On the statistical hypothesis testing and the computation of p-values, the null hypothesis is that the two variables (intensity and velocity) are uncorrelated. Thus this analysis result shows the correlation between the acceleration and intensity enhancement
(note that the negative correlation in the blue-shifted component means the positive acceleration). 

Further, by limiting the period of the flaring state of the maser features, we obtain the higher value of  cc is $-$0.87 with p-value of $6.5\times 10^{-15}$ (JD=2454500 to 2454650, see \S4.2) in the blue-shifted 
component, and cc is 0.78 with p-value of $2.4\times 10^{-10}$ (JD=2454610 to 2454770, the second flare, see \S4.2) in the red-shifted component, showing strong correlations during the flare states.
On the other hand, the systemic component show the strong correlation (cc = $-$0.79 with p-value of $9.9\times 10^{-41}$) regardless of flare states.

%%%%%\subsection{Correlation between the peak intensity and the optical light curve}
%%%%%
%%%%%The light curve of R Crt obtained by ASAS shows that the optical main period is estimated to be $P \sim$ 474 days 
%%%%%which is 3 times longer than the timescale of the period of 160 days suggested from AAVSO. 
%%%%%We carried out a cross correlation analysis between the maser intensity and the light curve for each velocity component by using the discrete correlation function
%%%%%introduced for unevenly sampled data (Edelson \& Krolik 1988), and show the results in Figure 6. 
%%%%%We used the bin size of the time lag of 10 days. 
%%%%%The magnitude of the ASAS data $m$ was transformed to the intensity $F$ in arbitary units by using below equation (Lekht et al. 2001),
%%%%%
%%%%%\begin{equation}
%%%%%\label{eq:fvis}
%%%%%  F = 2.512^{10.0-m}.
%%%%%\end{equation}
%%%%%
%%%%%
%%%%%Clear correlation peak can be found near both 470--490 and $-760$--$-810$ days in the blue-shifted and red-shifted components, respectively,
%%%%%and $-810$ day in the systemic component. Since a negative lag means that variation of the maser is delayed with respect to that
%%%%%of the light curve, it is likely that the lag time between the light curve and the maser intensity is 800 days. 
%%%%%However, the obtained lag time would be changed if the monitoring data increased, because our monitoring span of the water maser in
%%%%%R Crt is 1.3 year at present, which is much shorter than that of the optical light  with ASAS ($\sim 9$years). 

\section{Discussion}
\subsection{Acceleration in the maser shell}

It was known that R Crt emits the SiO maser with a single peak, and both the H$_2$O and OH masers with a double peak separated by 7 and 17 km s$^{-1}$, respectively (Jewell et al. 1991, Dickinson \& Chaisson 1973, G$\rm \acute o$mez Balboa et al. 1986,  Etoka et al. 2001. Etoka et al. 2003, Kim et al. 2010). 
This fact indicates that the materials are accelerated toward the outer region in the H$_2$O maser shell in the circumstellar envelope (Eliztur 1992). 
Averaged acceleration in the H$_2$O shell in R Crt can be estimated to be 0.4 km s$^{-1}$ yr$^{-1}$  
from the measured shell size and velocity range under the assumption of the simple thin shell model (I01). 
This value can be regarded as the averaged acceleration in the H$_2$O maser shell of R Crt, 
and is in good agreement with the typical acceleration obtained in the blue- and red-shifted components indicated from our observations. 
%%%%%As shown in \S 1, the H$_2$O masers in R Crt is expected to have 0.1 km s$^{-1}$ yr$^{-1}$ scale acceleration. 

In addition to linear acceleration, we found the velocity change between acceleration and deceleration, 
slightly correlated with the peak intensity of the maser for the first time (see \S 3.3). 
In order to explain the velocity variations in the stellar atmosphere of the AGB stars, the shock propagation model has been proposed (Wood 1979, Shintani et al. 2008). 
{  Based on this model, we consider a simple speculation of the origin of the change of acceleration. 
A shock wave is expected to be generated periodically by stellar pulsation and to propagate outward in the envelope. 
When the shock wave irrupts into the H$_2$O maser shell, it temporary causes both acceleration and intensity enhancement of the maser sources in the shell,
taking the collision pumping of the H$_2$O masers into consideration (Deguchi 1977).
Then after the shock wave passes through the H$_2$O maser shell, the accelerated maser source is considered to be become dark gradually.
Then, alternatively the maser sources in the shell behind the shock are found to be relatively bright again.
As a result, we observe apparent deceleration in the spectrum. 

However, since the actual situation seems to be more complex, it is not enough to consider only the shock propagation.
We encounter difficulties because (1) acceleration without the intensity variation was sometimes observed, and 
(2) the mechanism of the maser excitation or the intensity amplification is generally very complex because of beaming effect of maser emission.
Further, (3) VLBI mapping observations showed that an H$_2$O maser feature sometimes exhibits complex spatial and velocity structure, 
and the structure can easily change on the timescale of the order of a few months (e.g., Sudou et al. 2002), 
and (4) the observed velocity shift might not reflect the real bulk motion of the masers (e.g., Imai et al. 1997). 
We also cannot reject any other possibilities that the velocity shift is due to non-radial motion, such as a bipolar flow,
because the presence of the bipolar flow is implied by using the proper motion of the H$_2$O maser features (I01). 

It follows from  that we need to obtain further data sets of the H$_2$O spectrum including the proper motion with VLBI 
as well as the other maser lines in R Crt (e.g., SiO and OH).
Detailed 3-D motion and acceleration will be very useful to understand the shock propagation effect, 
and to incorporate the radiative transfer process in the envelope in order to understand dust-driven wind (e.g., Khouri et al. 2015). 
}
%%%Thus, the correlation between the velocity and the intensity can be one of the strong supporting factors for this model.
%%%Assuming the shock model, the lag time between optical light and the H$_2$O maser intensity indicates  
%%%the shock propagating time from the stellar surface to  the inner shell. 
%%%%In this case we have 
%%%%$v_{\rm shock} \sim 18 \left( \frac{r_0}{2\times10^{12}\rm m} \right)$  km/s. 
%%%%This implies the presence of supersonic flow in the envelope. 

\subsection{Relationship between the light curve and the maser variabilities}

{  The light curve of R Crt obtained by ASAS shows that the optical main period is estimated to be $P\sim474$ days, 
which is 3 times longer than the timescale of the period of 160 days suggested from AAVSO. 
We carried out a cross correlation analysis between the maser {  integrated} intensity and the light curve in Figure 6 
by using the discrete correlation function introduced for unevenly sampled data (Edelson \& Krolik 1988).
{  We found the maximum correlation between them at a lag time of 150 day with cc = 0.78. 
This corresponds to 0.3 $P$ for $P=474$ days. 
The secondary correlation peak can be seen at a lag time  of 480 day with cc = 0.71. %%%, about 330 days longer than the first lag time. 
%%%This is probably related to the separation time between the intensity peaks in the integrated total intensity (see Figure 2). 
In some supergiants, the similar correlation has been reported by using light curves with longer periods up to 20 years 
(e.g., Lekht et al. 2001, 2005, Rudnitskij 2005). They estimated the lag time to be 0.05 -- 0.5 $P$. 
On the other hand, no correlation was also reported from the data of a semi-regular variable RX Boo during 6 years (Winnberg et al. 2008). 
They suggested that the variation of the maser intensity could be due to the turbulence of the velocity coherence length of H$_2$O maser amplification.

We examined the same analysis between both the intensity and velocity for each maser component and the light curve, and 
we cannot found clear correlation between them with the lag time smaller than $P$ (see Figures 7 and 8). 
%%%This result might indicate the presence of the lag time longer than one pulsation period. 
%%%However, 
The tight correlation between the optical variation and the velocity variation of high-excitation CO lines formed much closer to 
the star than H$_2$O maser lines has been found in semi-regular variables on the 1 -- 10 km s$^{-1}$ scale (Lebzelter et al. 2000).
Since our present monitoring span of the water maser in R Crt of 1.3 year is much shorter than that of the optical light curve with ASAS
($\sim 9$ years), we  will discuss these correlations further if we would have more years of monitoring data of the water maser velocity.  
}
%%%%Thus, we cannot reject the possibility that the velocity shift is due to non-radial motion, rotation of the star, or bipolar flow.
%%%%%The origin of the large acceleration of the systemic component is also unclear (see \S 4.1). 

\section{Conclusion}

The H$_2$O maser spectra of a semi-regular variable R Crt were monitored in the radial velocity on the 0.1 km s$^{-1}$ scale
and peaked components in the radial velocity were exhaustively detected using our newly developed automatic method. 
The blue-shifted and red-shifted components exhibited switching between acceleration and deceleration. 
This velocity variation correlates with the intensity variation with the correlation coefficient of $\sim0.4$, 
in particular during the flaring state of the water maser with the correlation coefficient of $\sim 0.8-0.9$. 
%%%%\item {the intensity variation correlalte with the optical light curve with the lag time of 800 days, corresponding to 1.7 $P$.}
These facts can be explained by the idea that the shock propagation due to the pulsation make the maser sources brighten and accelerate simultaneously. 
{  However, this implication is not enough to explain the relationship between the maser intensity and velocity simply, 
because the mechanism of the maser excitation and acceleration is considered to be certainly complex. 
}
To support this idea more strongly, it is important to continue to frequently measure the velocity variation for 
longer periods by using occupied single dishes and to analyze such massive monitoring datasets by the automatic peak detection, 
as well as long-term optical monitoring.

\acknowledgments
We would like to thank the anonymous referees, who gave us many useful comments that improved the paper very much.
We also thank Takeru Suzuki and Yuki Yasuda for helpful conversations. 
This work is in part supported by JSPS KAKENHI (Grant-in-Aid for Scientific Research) \#25870322, \#16H02866 and JST PRESTO \#JPMJPR16N6.

\appendix

\section{Automatic Peak Detection Method}

Our developed peak detection method is based on a least squares fitting of a Gaussian basis function model. 
After the model is fitted to an observed maser spectrum, peaks are given by the center positions of the optimized basis functions. This section describes our method, which consist of the assumed model, optimization algorithm of model parameters and peak detection procedure.

\subsection{Peak model with Gaussian basis functions}

Let $x_n$ be the $n$-th observed value of velocity and  $y_n$ be the observed intensity at velocity $x_n$.
Because the observed $y$ can be negative due to observation noise, we replaced all negative values of $y$
with zeros.
All values of $x$ and $y$ are vectorized as $\bm{x}=(x_1,\dots,x_N)^\T$ and $\bm{y}=(y_1,\dots,y_N)^\T$,
where $N$ is the number of discretized velocity values. Then an observed spectrum consists of the pair of 
$\bm{x}$ and $\bm{y}$. 

As shown in figure 1(a), each peaked component has symmetric and exponentially decreasing components around the peak, which looks like a Gaussian function.  Our model approximates a peaked velocity component by a Gaussian basis function:
\ba
\phi(x; \mu,\sigma)
=
\exp\left( -\frac{(x-\mu)^2}{2\sigma^2} \right)
,
\ea
where $\mu$ is the center and $\sigma$ is the dispersion of the component.
And then the intensity of a maser spectrum at $x_n$ is approximated 
by a linear combination of Gaussian basis functions:
\ba
\hat{y}_n
=
\sum_{m=1}^M w_m \phi(x_n;\mu_m,\sigma_m)
=
\bm{w}^\T \bm{\phi}(x_n),~~ n=1,\dots,N
\ea
where $M$ is the number of basis functions, $\bm{w}=(w_1,\dots,w_M)^\T$ are parameters of a non-negative weight vector and 
$\bm{\phi}(x_n) = \Big( \phi(x_n;\mu_1,\sigma_1),\dots, \phi(x_n;\mu_M,\sigma_M) \Big)^\T$
is a vector consisting of $M$ Gaussian basis functions.

Each basis function $\phi_m$ has three parameters, {\it i.e.} weight (height) $w_m$, center location $\mu_m$,  and dispersion (variance) $\sigma^2_m$, and then the model totally have $3\times M$ parameters; $\bm{\theta}=(\bm{w},\bm{\mu},\bm{\sigma})$, to be optimized based on least squares fitting to an observed spectrum.
$M$ is also a parameter but this cannot be optimized from an observed spectrum because increasing $M$ causes overfitting. We first examined all the spectra and then we manually set $M=10$ in our data analysis. Roughly choosing the value of $M$ does not cause an incorrect result because the weight parameters of unnecessary components tend to be small values and some of the components located close each other were removed in our post-processing procedure.

\subsection{Parameter optimization algorithm}

A least squares fitting is a parameter optimization based on minimizing the squared error
between observed values and estimated values:
\ba
J(\bm{\theta})
=
\frac{1}{2}\sum_{n=1}^N\left\{ y_n - \bm{w}^\T\bm{\phi}(x_n) \right\}^2.
\ea
In our problem setting, the optimization problem for all parameters $\bm{\theta}$ is non-convex, where there are many local optima.
Our algorithm first initialize parameters and then iteratively optimizes each parameter one-by-one with certain fixed parameters until convergence.
This procedure is repeatedly implemented from different initial parameters. Among optimization results from different initializations, the best model minimizing $J(\bm{\theta})$ is chosen as the final fitting result. The reminder of this section presents an optimization of each parameter,
an initialization procedure, and a post-processing procedure to remove redundant components.\\

\noindent{ Optimization of $\bm{w}$:}

With fixed $\bm{\mu}$ and $\bm{\sigma}$, the error function $J(\bm{w})$ is a quadratic convex function regarding to $\bm{w}$. 
Thus using the equation $\frac{\partial}{\partial\bm{w}}J(\bm{w})=0$, 
the optimal $\bm{w}$ can be analytically computed by
\ba
\bm{w}=(\bm{\Phi}^\T\bm{\Phi})^{-1}\bm{\Phi}^\T\bm{y},
\label{eq:update_w}
\ea
where
\ba
\bm{\Phi} = 
\left(
\begin{array}{cccc}
		\phi(x_1;\mu_1,\sigma_1) & \phi(x_1;\mu_2,\sigma_2) & \ldots & \phi(x_1;\mu_M,\sigma_M) \\
		\phi(x_2;\mu_1,\sigma_1) & \phi(x_2;\mu_2,\sigma_2) & \ldots & \phi(x_2;\mu_M,\sigma_M) \\
		\vdots & \vdots & \ddots & \vdots \\
		\phi(x_N;\mu_1,\sigma_1) & \phi(x_N;\mu_2,\sigma_2) & \ldots & \phi(x_N;\mu_M,\sigma_M) 
\end{array}
\right).
\ea
After updating $\bm{w}$ using equation (\ref{eq:update_w}), the elements of $\bm{w}$ can be negative. 
Thus, our algorithm replaces all negative values in $\bm{w}$ with zeros after the update (\ref{eq:update_w}).
\\

\noindent{ Optimization of $\bm{\mu}$ and $\bm{\sigma}$:}

The error functions of $\bm{\mu}$ and $\bm{\sigma}$ are not quadratic convex functions, such that the optimal parameter 
cannot be computed analytically. Our method used a gradient descend approach to optimize these parameters, 
which is given by
\ba
\bm{\mu} &\leftarrow& \bm{\mu} - \eta \frac{\partial J(\bm{\theta})}{\partial \bm{\mu}}
\label{eq:update_mu}
\\
\bm{\sigma} &\leftarrow& \bm{\sigma} - \eta \frac{\partial J(\bm{\theta})}{\partial \bm{\sigma}},
\label{eq:update_sigma}
.
\ea
$\eta$ is s step size to be set to a small enough value ($\eta = 10^{-3}$ in our experiment) and
\ba
\frac{\partial J(\bm{\theta})}{\partial \bm{\mu}} &=&
\left( \frac{\partial J(\bm{\theta})}{\partial \mu_1}, \dots, \frac{\partial J(\bm{\theta})}{\partial \mu_M}\right)^\T
,
\\
\frac{\partial J(\bm{\theta})}{\partial \bm{\sigma}} &=&
\left(\frac{\partial J(\bm{\theta})}{\partial \sigma_1}, \dots, \frac{\partial J(\bm{\theta})}{\partial \sigma_M}\right)^\T 
.
\ea
In the update rule, the first derivatives are computed by
\begin{eqnarray}
\frac{\partial J}{\partial \mu_m} 
& = & \sum_{n=1}^N \left\{y_n-\sum_{j=1}^M{w_j}{\exp\left(-\frac{(x_{n}-\mu_j)^2}{2\sigma_j^2}\right)}\right\}\non \\
& ~ & \left\{-\frac{w_m(x_{n}-\mu_m)}{\sigma_m^2}{\exp\left(-\frac{(x_{n}-\mu_m)^2}{2\sigma_m^2}\right)}\right\} \non \\
\\
\frac{\partial J}{\partial \sigma_m} 
& = & \sum_{n=1}^N \left\{y_n-\sum_{j=1}^M{w_j}{\exp\left(-\frac{(x_{n}-\mu_j)^2}{2\sigma_j^2}\right)}\right\} \non \\
& ~ & \left\{-\frac{w_m(x_{n}-\mu_m)^2}{\sigma_m^3}{\exp\left(-\frac{(x_{n}-\mu_m)^2}{2\sigma_m^2}\right)}\right\}
. \non\\
\end{eqnarray}
Our optimization updates these parameters $\bm{w}$, $\bm{\mu}$ and $\bm{\sigma}$ alternately until convergence.

In this optimization, initializing the center parameter $\bm{\mu}$
is essentially important to reach to a good local optimum for an appropriate fitting result. 
To choose an appropriate initialization of $\bm{\mu}$,
{  we implemented the following procedure: we first set the threshold value $h_{\rm noise}$  ($h_{\rm noise}$ = 0.1 Jy/beam in our experiment)}
to remove observation noise, and then listed all position $x_n$ whose output $y_n$ is greater than $h_{\rm noise}$. 
Next we randomly select a position $x$ from the list and set to an initial value of center $\mu_1$. 
{  Then the selected value and the values whose distance from the selected value 
is less than $h_{\rm dist}$ ($h_{\rm dist}=0.1$ kms$^{-1}$ in our experiment) are removed from the list. }
The next initial value of $\mu_2$ is then randomly chosen from the list.
This procedure is iterated until the number of selected centers reaches to $M$. 
Due to the threshold $h_{\rm dist}$, the list can be empty after some initial positions are chosen. 
In this case, the value of $h_{\rm dist}$ decreased to the half, {\it i.e.} $h_{\rm dist}/2$, 
and the initialization procedure continues.

After optimizing the parameters, the centers of the basis functions can be similar. As a post-processing procedure, 
our method chooses only the basis function with larger weight parameter $w$ from close basis functions 
(for which the distance between the centers is less than 0.5) and removes the other basis function. 
Then the parameters are optimized again. 
Even after our optimization, there can be several small components because of observation noise.
{  Since the observed spectrum features with low signal-to-noise ratio hard to be decomposed correctly due to
blending effect of complex features, we removed components with   {$w_m\le w_{\rm thre}$}, %%%This leads the fitting error of the
which each spectral feature to be almost negligible compared with the observation errors. }
%%%\red {In our analysis, we set $w_{\rm noise} = 3 h_{\rm noise}$  empirically. }
Then our method outputs the center of the remaining basis functions as peak positions in the observed spectrum. 

%%%\newpage

%%%%%%%%%%%%%%%%%%%%%%%%%%%%%%%%%%%%%%%%%%%%%%%%%%%%%%%%%%%%%%%%%%%%%%%%%%%%%%%%%%%%%%%%%%%%%%%%%%
%%%%%%%%Figure
%%%%%%%%%%%%%%%%%%%%%%%%%%%%%%%%%%%%%%%%%%%%%%%%%%%%%%%%%%%%%%%%%%%%%%%%%%%%%%%%%%%%%%%%%%%%%%%%%%
\onecolumn
%%%%%%%%%%%%%%%%%%%%figure 1
\begin{figure}
 \begin{center}
 (a) Observed H$_2$O maser spectrum\hspace{20mm}\\
  \includegraphics[width=10cm]{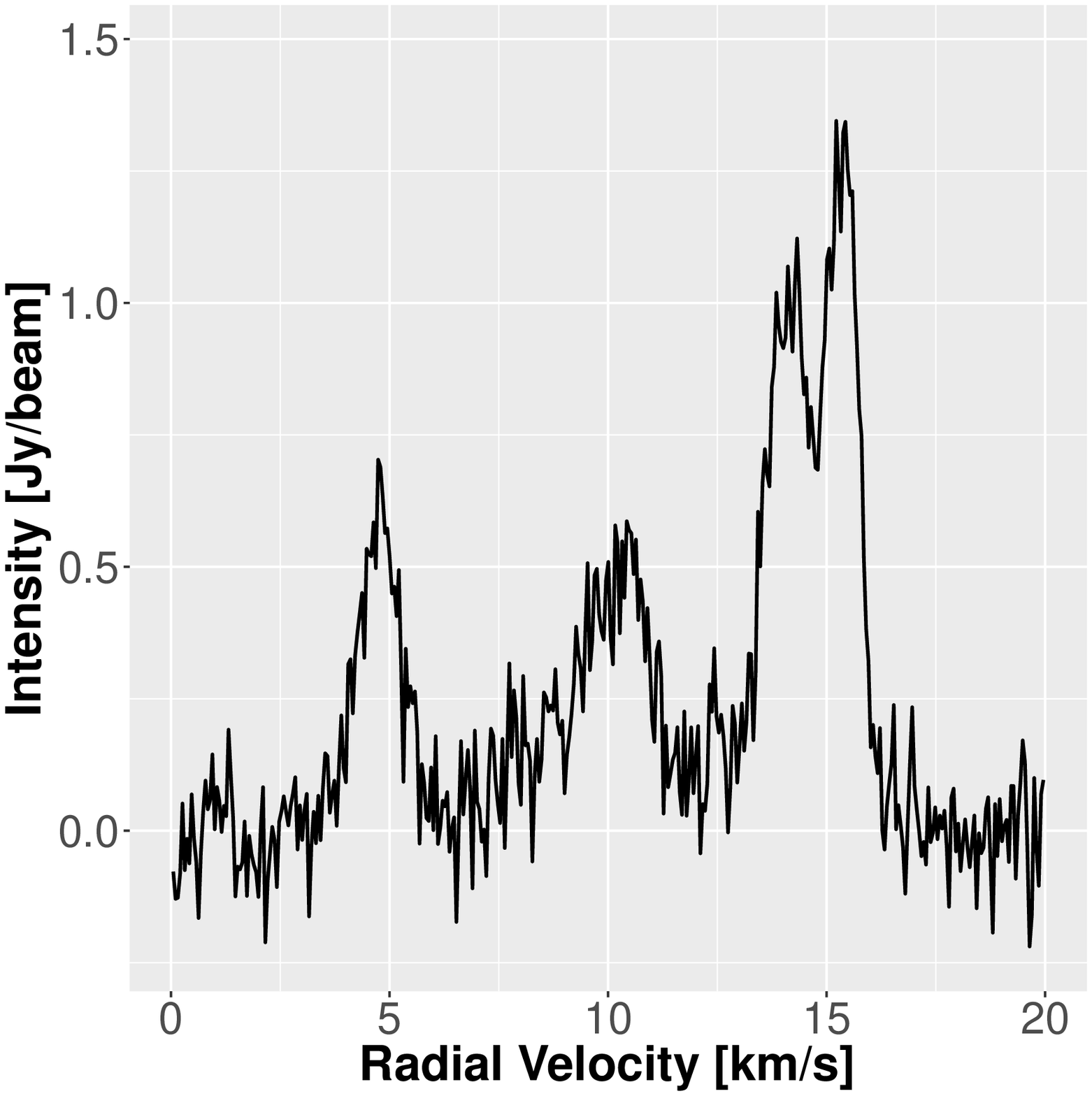}
  \\
 (b) Peak fitting result with Gaussian basis functions \hspace{20mm}\\ 
  \includegraphics[width=10cm]{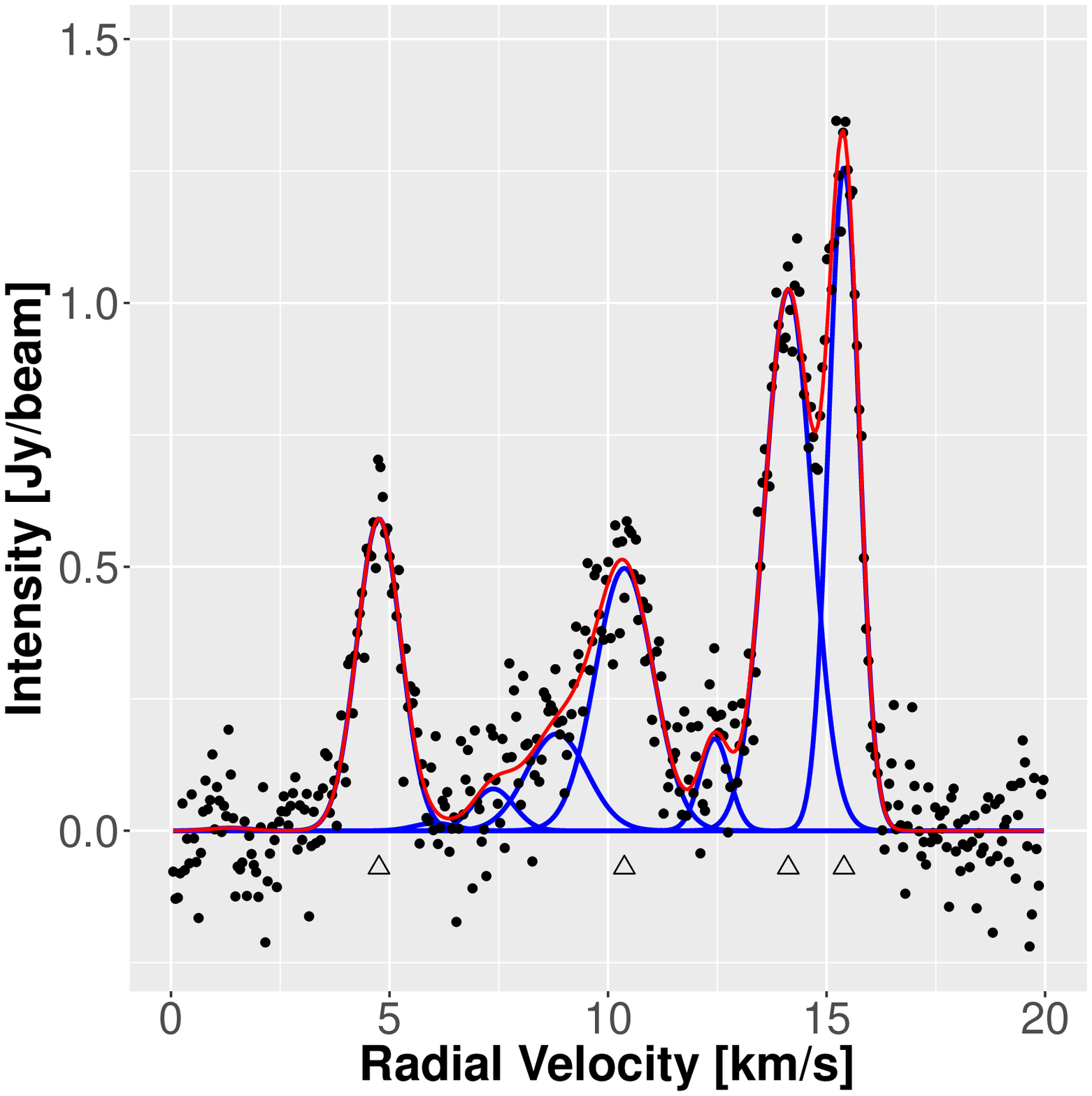}
 \end{center}
 \caption{Typical spectra of H$_2$O maser emission in R Crt (observed on January 13, 2008) and fitting result. 
 (a) Observed maser spectrum, in which the systemic velocity is 10.8 kms$^{-1}$. 
 (b) Fitting result. Circles: observed data, red line: fitting result by equation (1), triangles: detected peaks 
 (only significant components, $w_m\ge 0.3$) Jy/beam.
 }
 \label{fig1}
\end{figure}

%%%%%%%%%%%%%%%%%%%%figure 2
\begin{figure}
 \begin{center}
  \includegraphics[width=15cm]{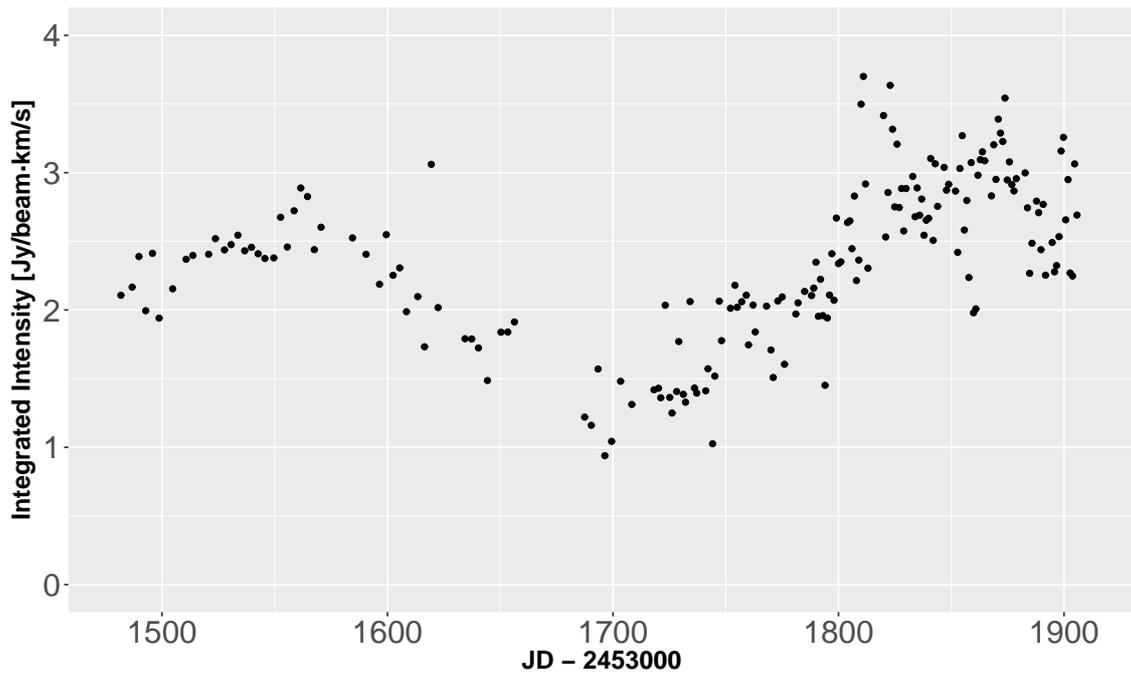}
 \end{center}
    \vspace{50mm}
 \caption{Time evolution of the integrated intensity. }\label{fig2}
\end{figure}

%%%%%%%%%%%%%%%%%%%%figure 3
\begin{figure}
 \begin{center}
  \includegraphics[width=15cm]{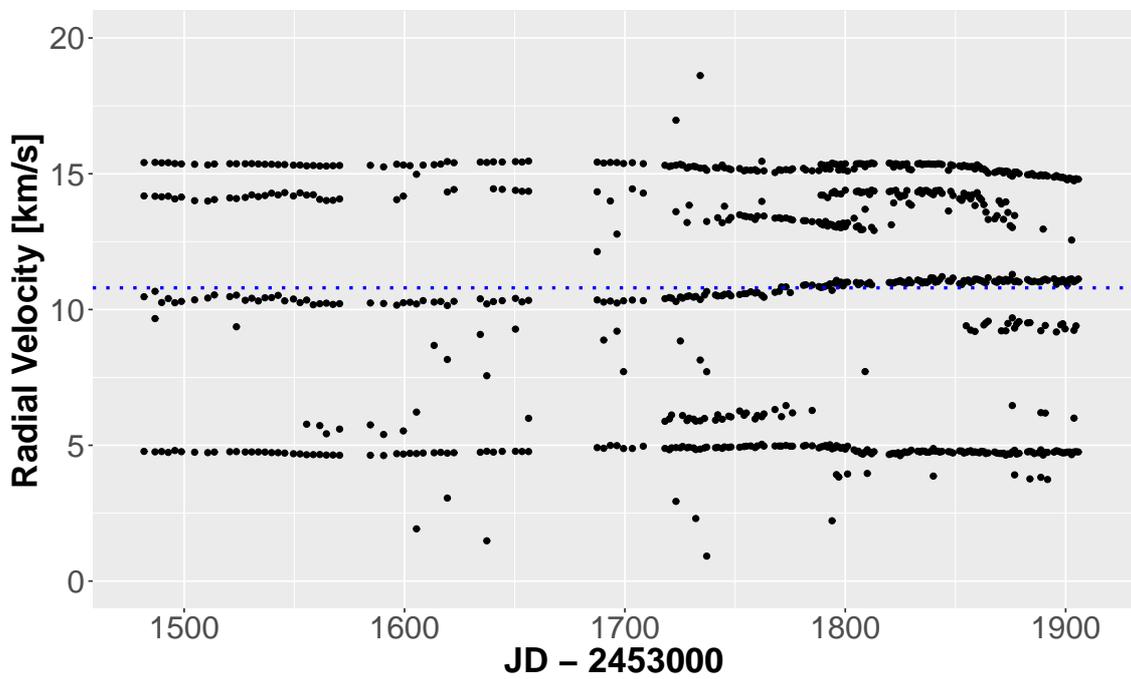}
 \end{center}
 \caption{Time evolution of the fitted peak radial velocity covering the whole velocity range. The blue dotted line indicates the stellar velocity (10.8 km s$^{-1}$). }\label{fig2}
\end{figure}

%%%%%%%%%%%%%%%%%%%%figure 4
\begin{figure}
\begin{minipage}{0.5\hsize}
  \begin{center}
  (a) blue-shifted component\\
  %%%\vspace{-5mm}
   \includegraphics[width=6cm,angle=270]{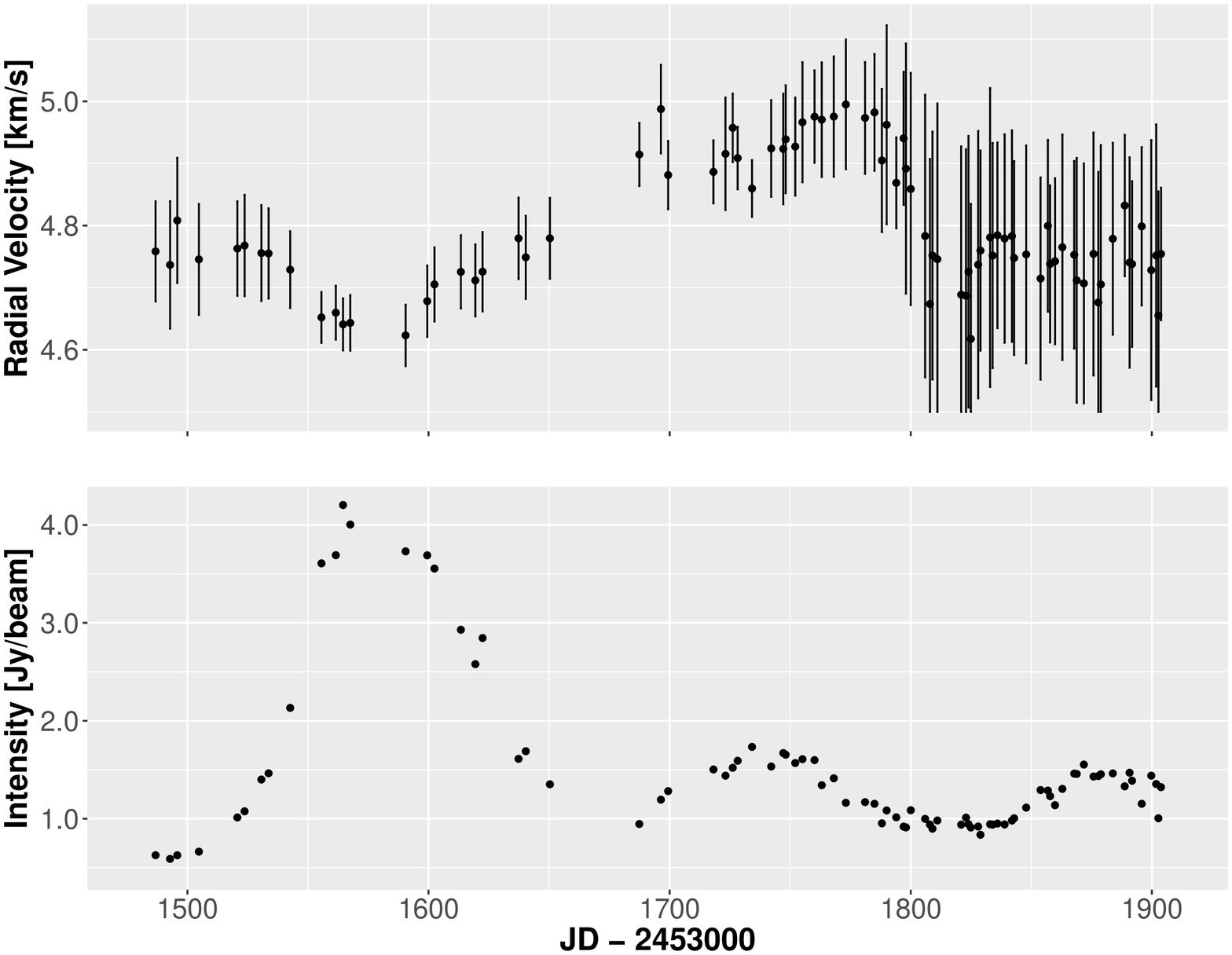}
  \end{center}
  \vspace{10mm}
 \end{minipage}
 \begin{minipage}{0.5\hsize}
  \begin{center}
  (b) systemic component\\
  %%%\vspace{-5mm}
   \includegraphics[width=6cm,angle=270]{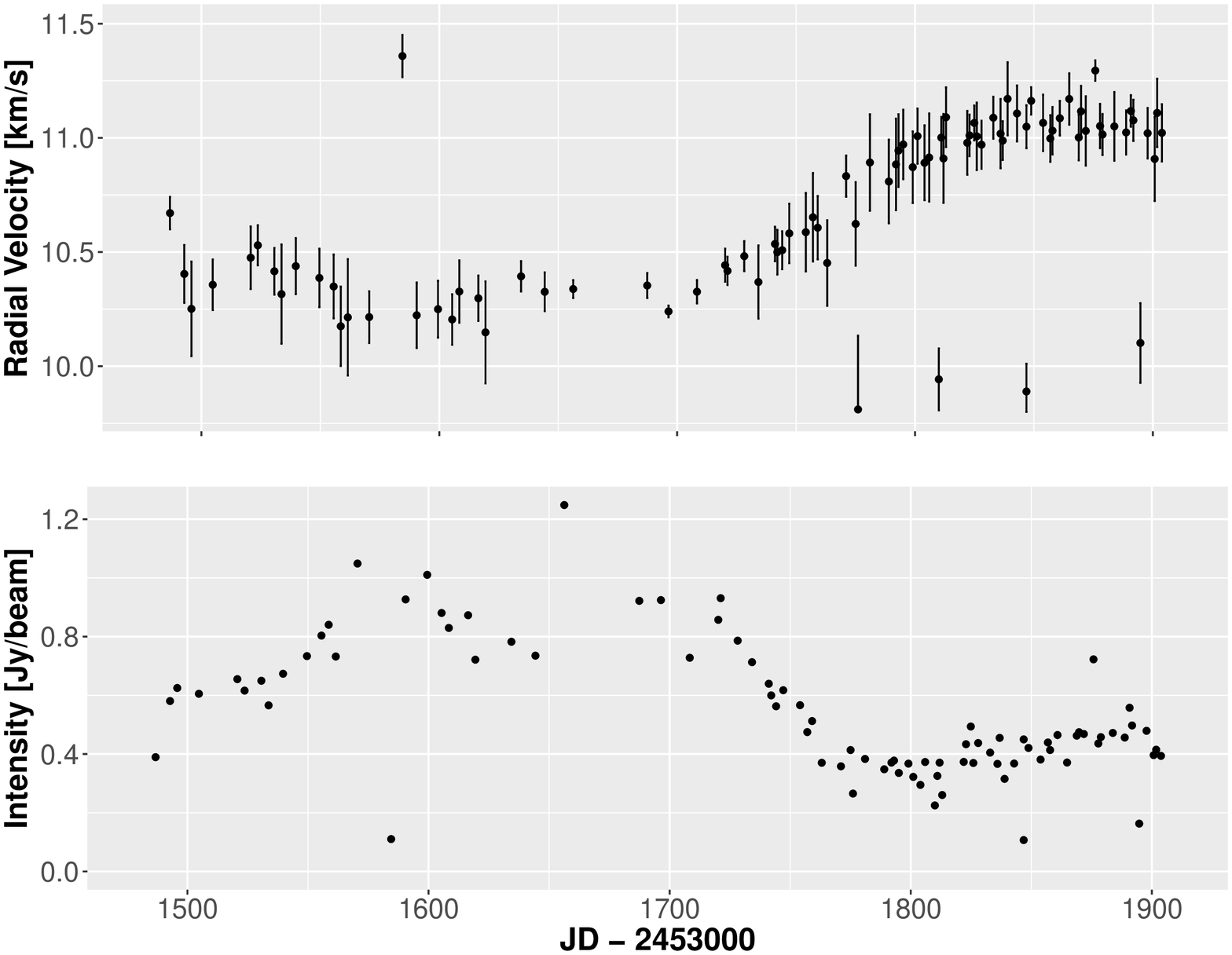}
  \end{center}
  \vspace{10mm}
 \end{minipage}
 \begin{minipage}[t]{0.5\hsize}
  \begin{center}
  (c) red-shifted component\\
  %%%\vspace{-5mm}
   \includegraphics[width=6cm,angle=270]{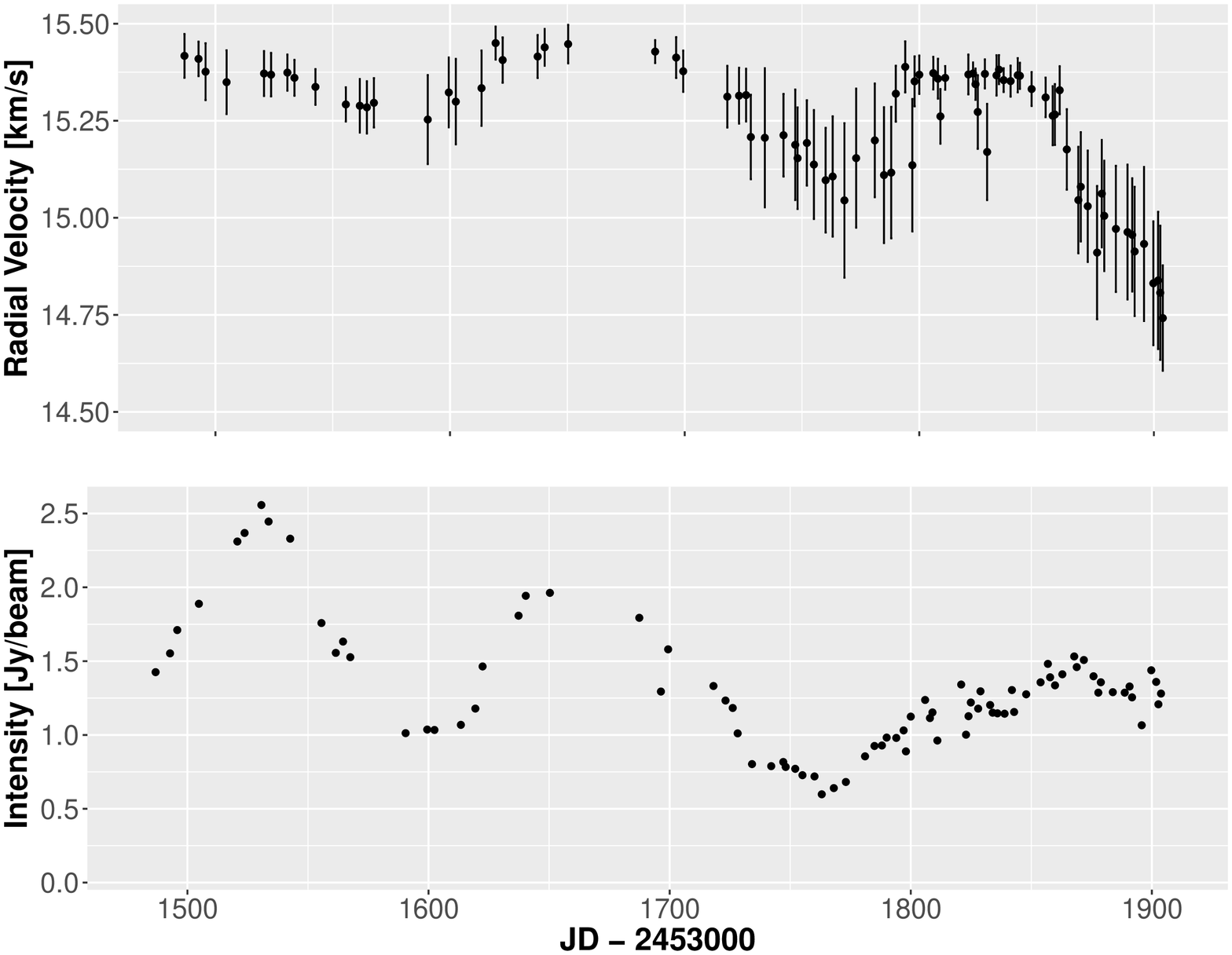}
  \end{center}
 \end{minipage}
 \begin{minipage}[t]{0.5\hsize}
  \begin{center}
  (d) the optical light curve \\
  %%%\vspace{-5mm}
   \includegraphics[width=6cm,angle=270]{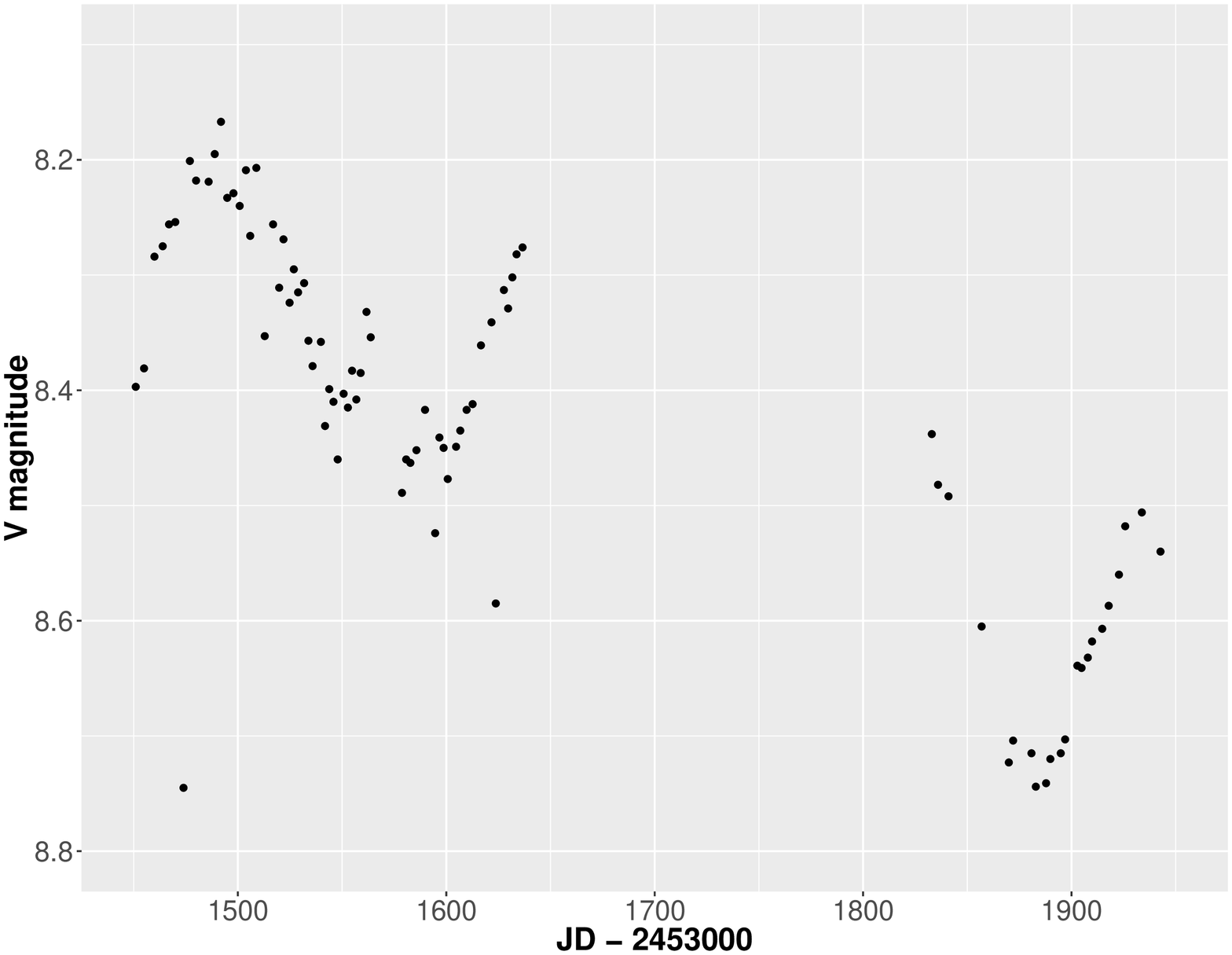}
  \end{center}
 \end{minipage}
  \caption{Time evolution of (a) blue-shifted component, (b) systemic component, and (c) red-shifted component. 
 %%%The top panel shows the radial velocity, the middle panel shows the peak intensity of H$_2$O masers, and
 %%%the bottom panel shows the velocity width. (d) optical light by ASAS.}\label{fig3}
 The top panel shows the radial velocity (circles) and the velocity width (bars), and the bottom panel shows the peak intensity of H$_2$O masers.
  (d) optical light by ASAS.}\label{fig3}
\end{figure}  

%%%%%%%%%%%%%%%%%%%%figure 4old
%%%\begin{figure}
%%% \begin{center}
 %%% \includegraphics[width=15cm]{acc-all.eps}
%%% \end{center}
%%% \caption{acceleration}\label{fig4}
%%%\end{figure}

%%%%%%%%%%%%%%%%%%%%figure 4
%%%%\begin{figure}
%%%% \begin{center}
%%%%  \includegraphics[width=10cm]{fig3d.eps}
%%%% \end{center}
%%%% \caption{The optical light curve of R Crt obtained by ASAS. }\label{fig4}
%%%%\end{figure}

%%%%%%%%%%%%%%%%%%%%figure 5
\begin{figure}
\begin{minipage}{0.5\hsize}
  \begin{center}
  (a) blue-shifted component\\
  %%%\vspace{-5mm}
   \includegraphics[width=6cm,angle=0]{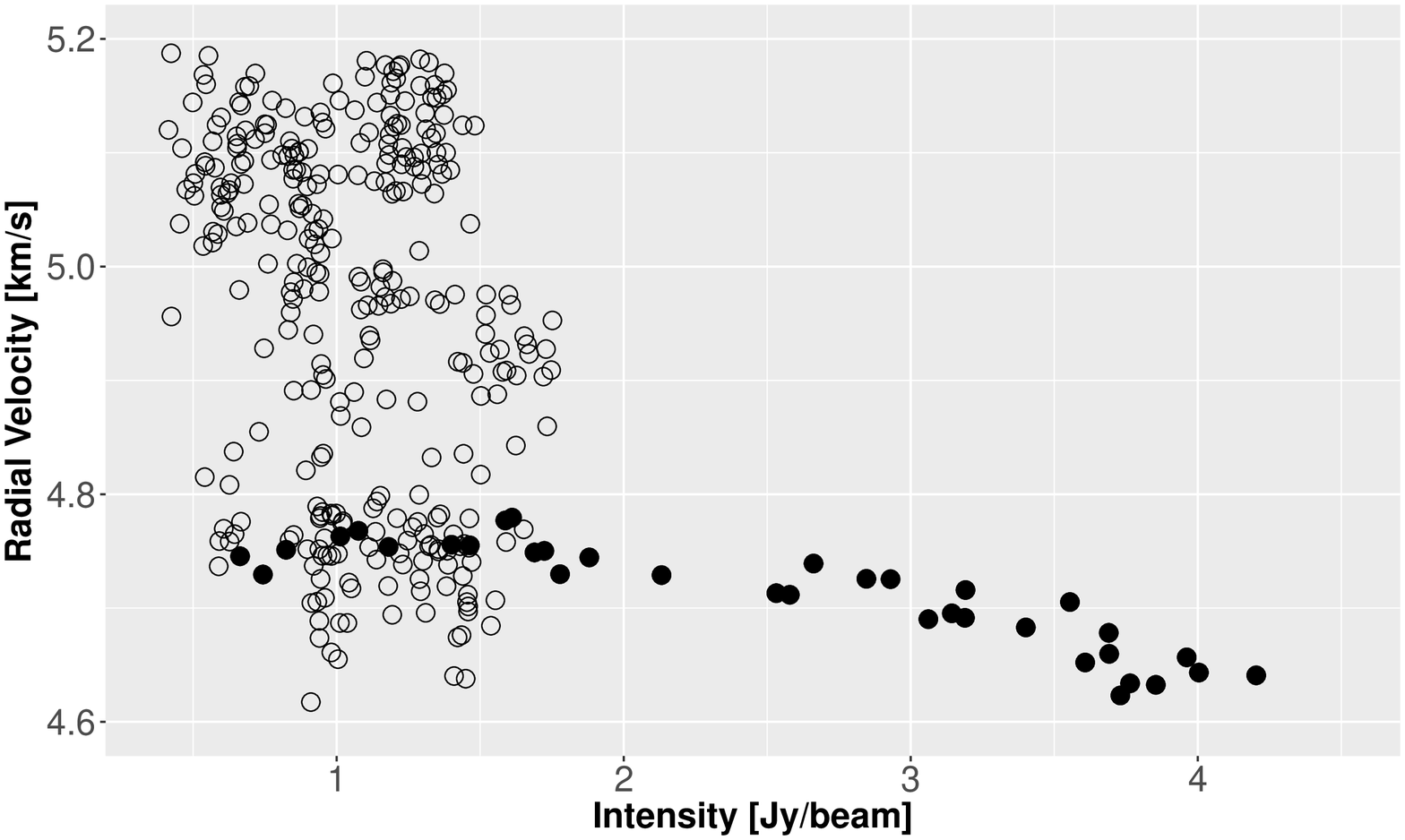}
  \end{center}
  \vspace{10mm}
 \end{minipage}
 \begin{minipage}{0.5\hsize}
  \begin{center}
  (b) systemic component\\
  %%%\vspace{-5mm}
   \includegraphics[width=6cm,angle=0]{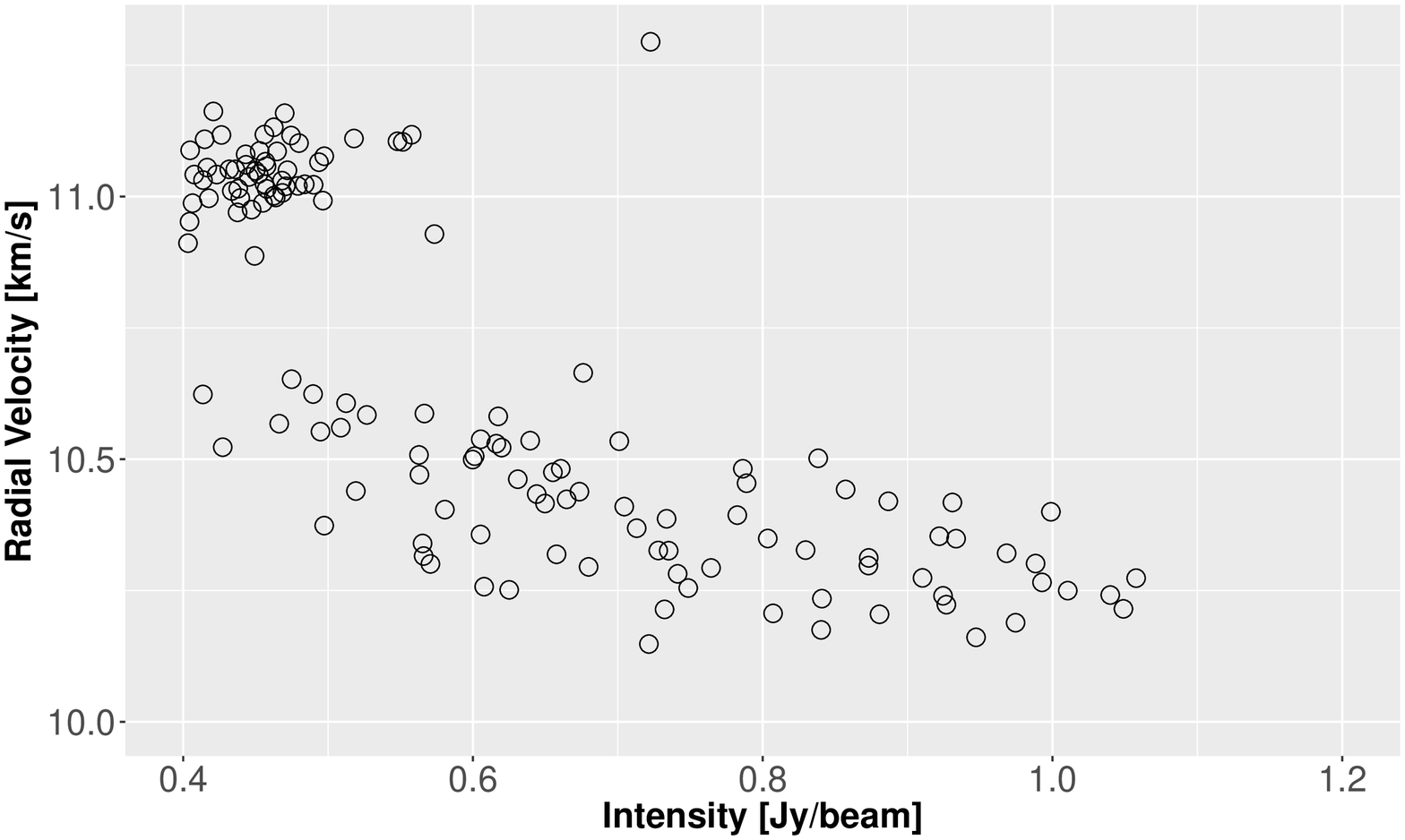}
  \end{center}
  \vspace{10mm}
 \end{minipage}
 \begin{minipage}[t]{0.5\hsize}
  \begin{center}
  (c) red-shifted component\\
  %%%\vspace{-5mm}
   \includegraphics[width=6cm,angle=0]{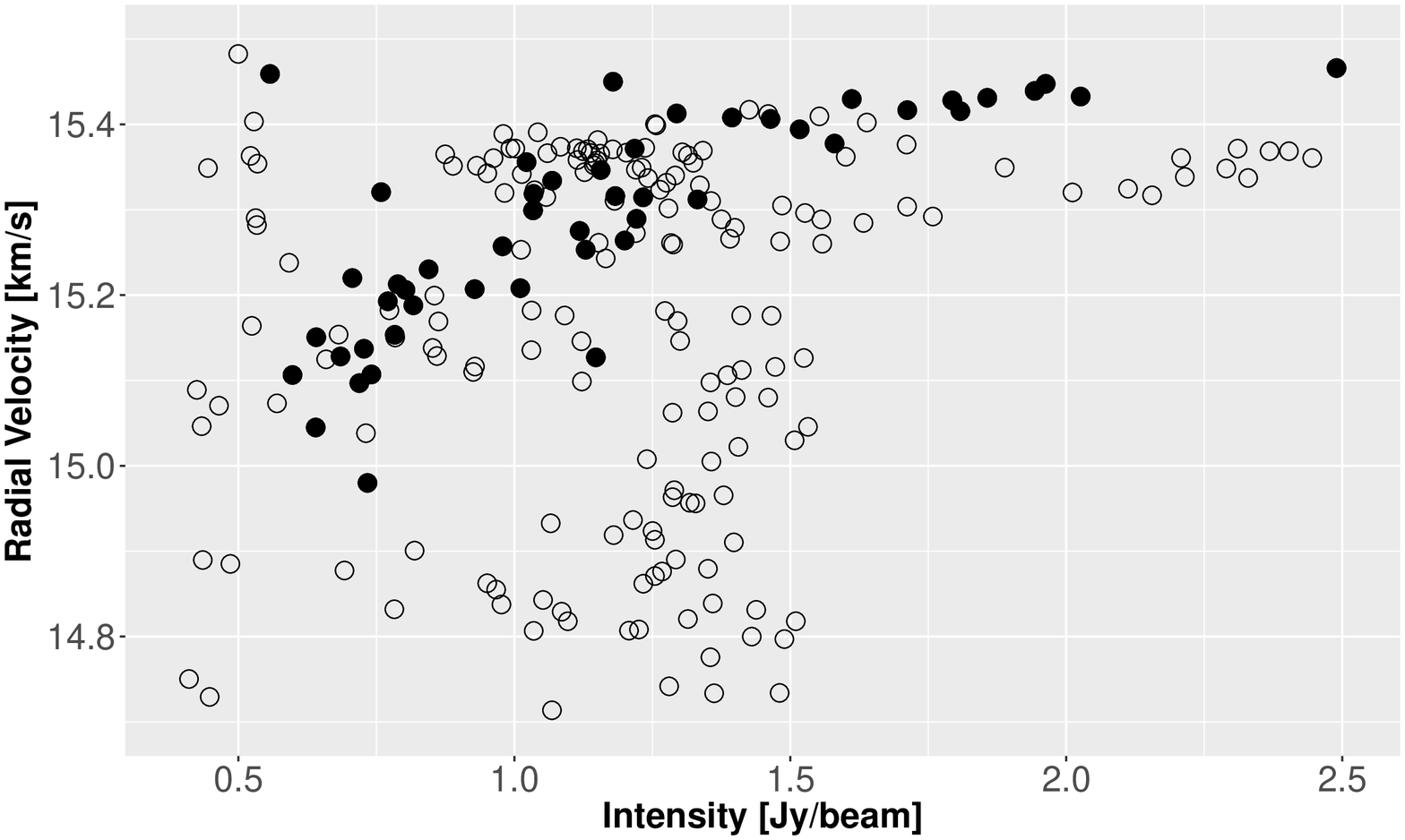}
  \end{center}
 \end{minipage}
  \caption{The correlation between the radial velocity and the intensity of the water maser. (a) blue-shifted component, (b) systemic component, and (c) red-shifted component.
   The open circles indicate the data in the whole observation period. 
   The filled circles indicate the data  obtained in the period of the flaring state between (a) JD=2454500 and 2454650, and (c) JD=2454600 and 2454770.}\label{fig4}
\end{figure}

%%%%%%%%%%%%%%%%%figure 6
\begin{figure}
 \begin{center}
  \includegraphics[width=10cm]{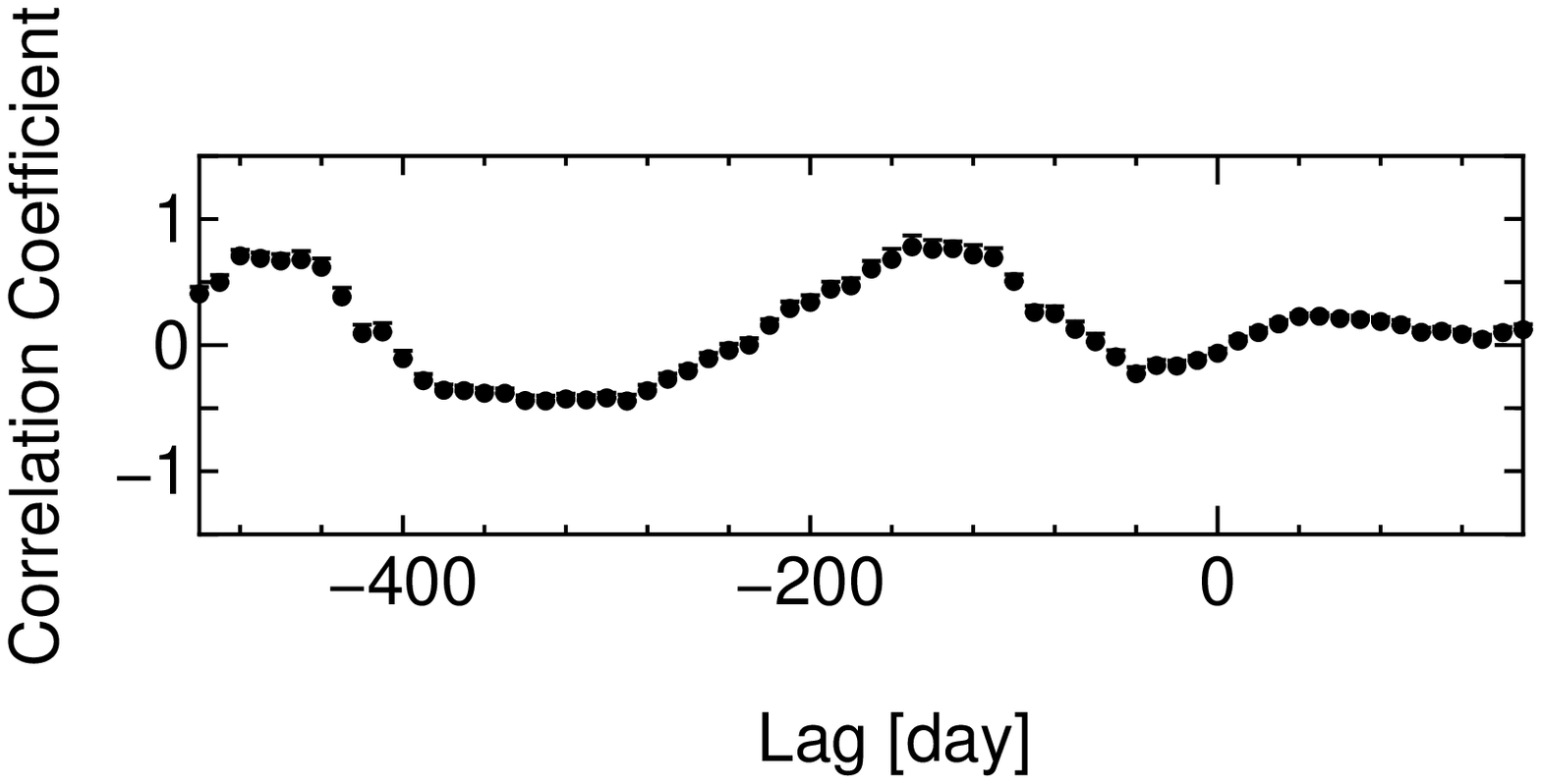}
 \end{center}
   \vspace{50mm}
 \caption{The discrete correlation function between the optical light curve and the integrated maser intensity. 
   A negative lag means that variation of the maser is delayed with respect to that of the light curve. }\label{fig5}
\end{figure}

%%%%%%%%%%%%%%%%%figure 7
\begin{figure}
 \begin{center}
  \includegraphics[width=10cm]{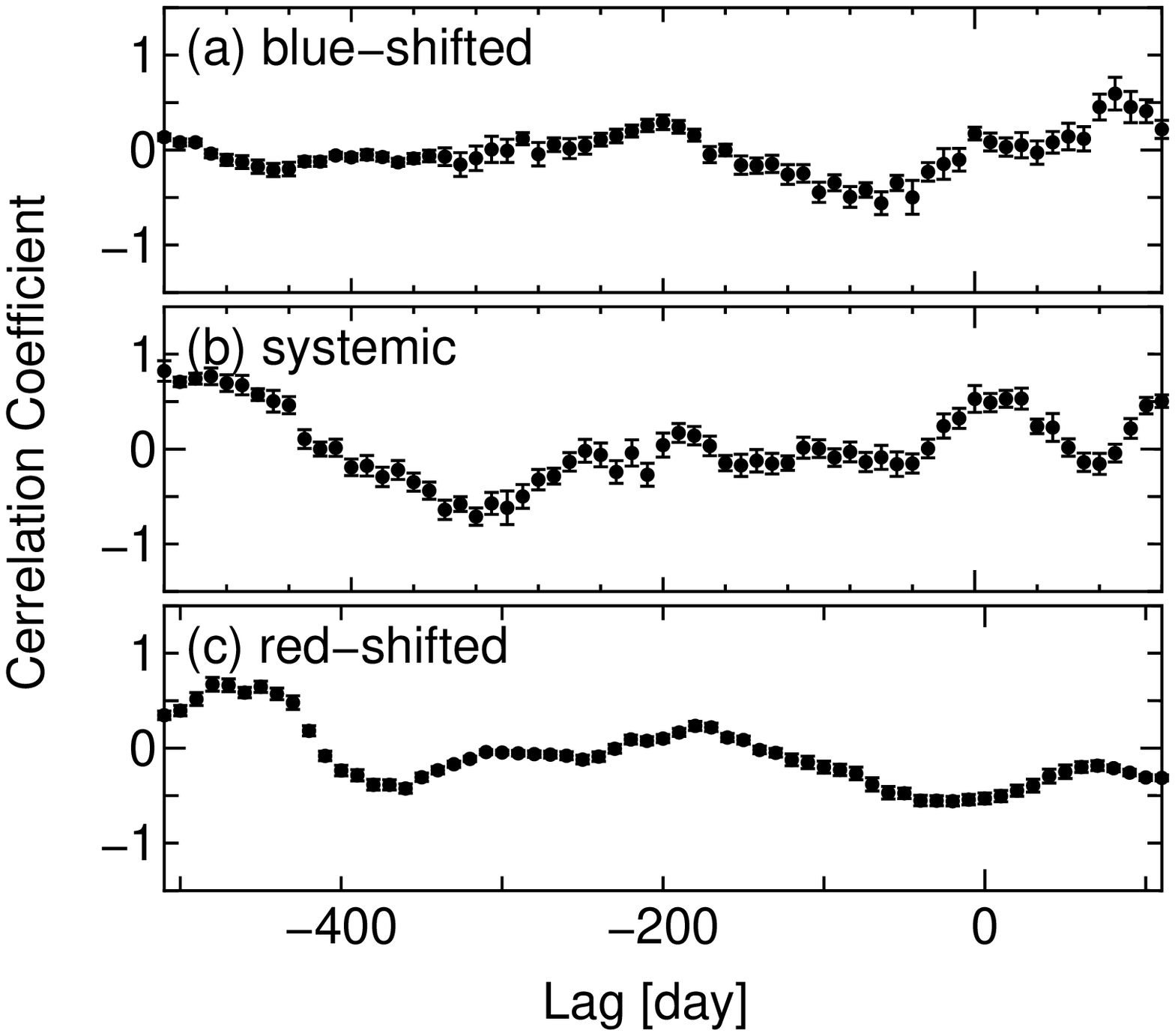}
 \end{center}
   \vspace{100mm}
 \caption{The discrete correlation function between the optical light curve and the maser intensity. 
   A negative lag means that variation of the maser is delayed with respect to that of the light curve. }\label{fig5}
\end{figure}

%%%%%%%%%%%%%%%%%figure 8
\begin{figure}
 \begin{center}
  \includegraphics[width=10cm]{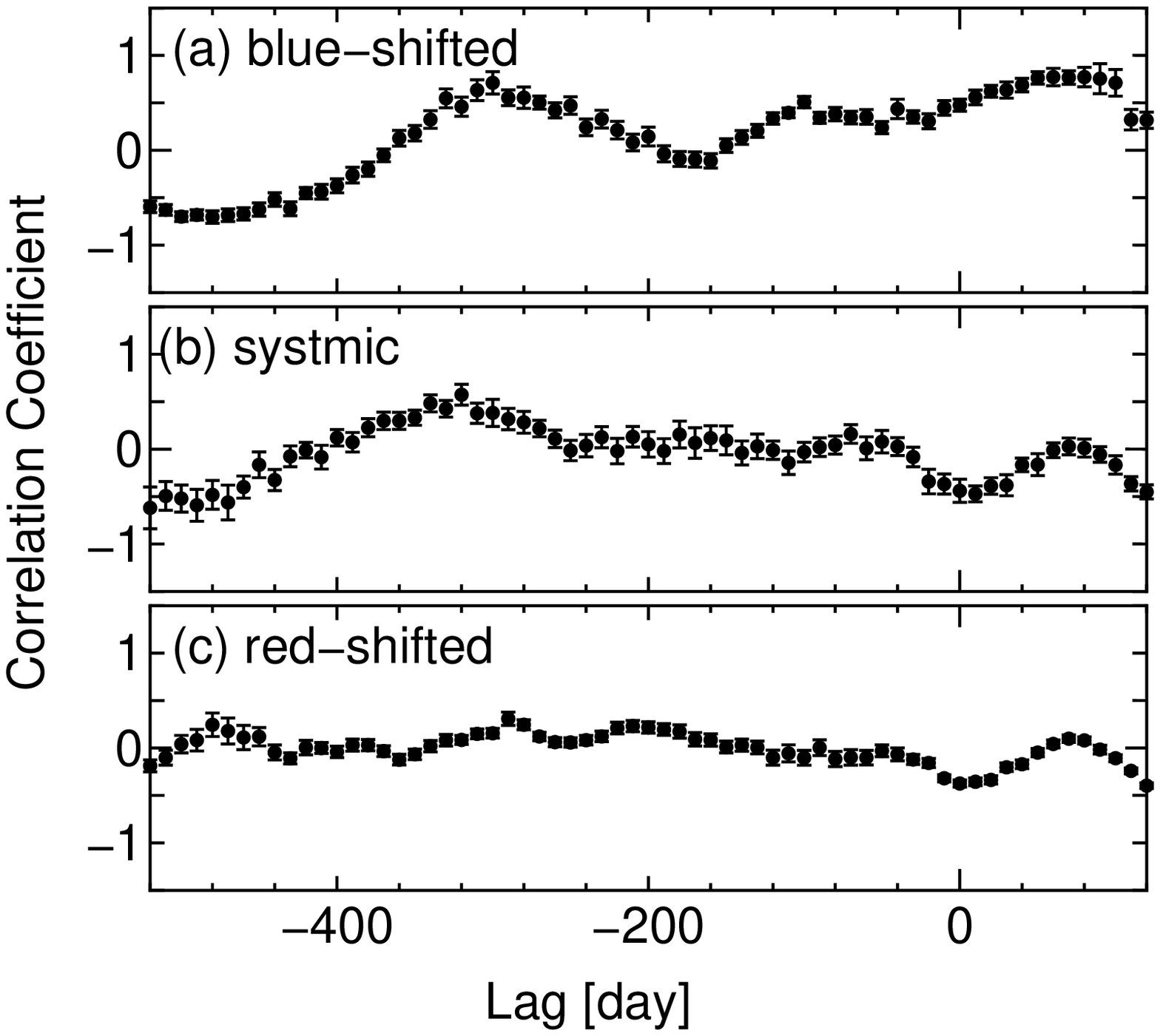}
 \end{center}
    \vspace{50mm}
 \caption{The discrete correlation function between the optical light curve and the maser velocity. 
   A negative lag means that variation of the maser is delayed with respect to that of the light curve. }\label{fig6}
\end{figure}

\end{document}